# INSTRUMENTATION AND ITS INTERACTION WITH THE SECONDARY BEAM FOR THE FERMILAB MUON CAMPUS *

D. Stratakis[†], B. Drendel, M. J. Syphers[1] Fermi National Accelerator Laboratory, Batavia IL, USA
[1]also at Northern Illinois University, DeKalb IL, USA

*Abstract*

The Fermilab Muon Campus will host the Muon g-2 experiment - a world class experiment dedicated to the search for signals of new physics. Strict demands are placed on beam diagnostics in order to ensure delivery of high quality beams to the storage ring with minimal losses. In this study, we briefly describe the available secondary beam diagnostics for the Fermilab Muon Campus. Then, with the aid of numerical simulations we detail their interaction with the secondary beam. Finally, we compare our results against theoretical findings.

## INTRODUCTION

The Muon g-2 Experiment, at Fermilab [1], will measure the muon anomalous magnetic moment, $\alpha_\mu$ to unprecedented precision: 0.14 parts per million. A sequence of beamlines that are part of the Fermilab Muon Campus [2] have been designed to deliver the highest possible quality muon beam for the experiment. Figure 1 displays a schematic layout of the Fermilab Muon Campus. Two groups eight, 120 ns bunches of $10^{12}$ protons each are directed to a target station. Further downstream, a secondary beam of positively-charged particles with a momentum of 3.1 GeV/c (± 10%) will be selected using a bending magnet. The beam leaving the target station will travel through the M2 and M3 lines which are designed to capture as many muons from pion decay as possible. The beam will then be injected into the Delivery Ring (DR) wherein all pions will decay into muons, and the muons will separate in time from the heavier protons. The muon beam will be extracted into another combination of beamlines (known as M4 and M5 lines) which end just upstream of the entrance of the Muon g-2 Experiment storage ring.

Beam monitoring along the Muon Campus [3] can be divided into four different zones, each with different instrumentation schemes. In particular, high-intensity proton beam will be monitored with toroids, beam position monitors (BPMs) and beam loss monitors (BLMs). On the other hand, low-intensity secondary and proton-only secondary beam will be monitored with ion chambers, BLMs and secondary emission monitors (SEMs). Muon-only beam will be monitored with ion chambers and proportional wire chambers (PWCs). In this paper, we will overview the interaction of the secondary beam with the aforementioned instrumentation. Since our prime interest is the secondary beam we will give more emphasis on ion chambers and PWCs monitors.

## MUON CAMPUS SECONDARY BEAM DIAGNOSTICS

Ion chambers will become the primary beam-intensity measurement device for the muon beam. They are relatively inexpensive devices that can measure beam intensities with an accuracy of ±5% with as little as $10^5$ particles. For the Muon g-2 Experiment operations, ion chambers will be implemented in the M2-M3 lines, the Delivery Ring (DR) and the M4-M5 lines. One advantage is that these ion chambers will be installed in a vacuum-can with motor controls to allow them to be pulled out of the beam.

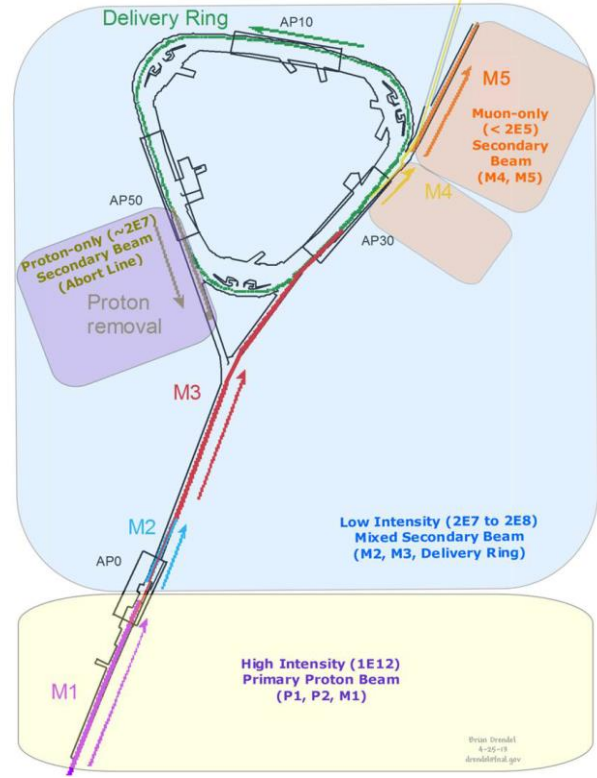

Figure 1: Schematic illustration of the Muon Campus beamlines. Devices will be distributed along the Muon Campus based on beam intensity. More details can be found in Ref. 3.

Figure 2 shows the ion chamber design. Each ion chamber consists of one signal foil interleaved between 24 high-voltage foils. The foils are sealed in an aluminum chamber continuously purged with an 80% argon - 20% carbon dioxide gas mix. The ion chamber will be installed inside of an anti-vacuum chamber with two titanium vacuum windows to provide a barrier between the gas needed for the ion chamber and the beamline vacuum. The entire anti-vacuum chamber would be mounted inside of a vacuum can that is common to beam tube vacuum. The ion chamber will be on a motorized drive that would allow it to be moved in or out of the path of beam. This may reduce beam degradation from multi-scattering effects.

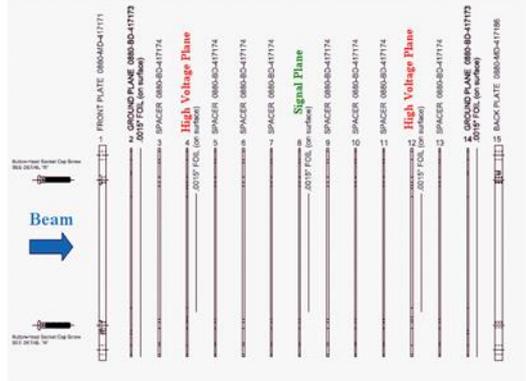

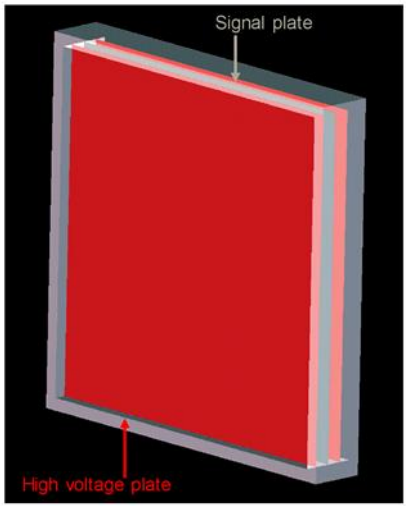

Figure 2: Schematic illustration of the ion chamber detector. The bottom figure shows the simulation model.

Beam profiles of muons will be measured using PWCs. PWCs are sensitive, since they have the capability to measure beam intensities down to the $10^3$ particle range. When mounted inside refurbished Switchyard bayonet vacuum cans, the PWCs can be pulled out of the beam path when not in use. This eliminates the need for permanent vacuum windows and vacuum bypasses. The PWC has two planes of signal wires, one plane for horizontal and one for vertical. There are 48 signal wires in each plane which are 10 μm diameter gold-plated tungsten and can be configured with either 1 mm or 2 mm spacing. The wire planes are sandwiched between Aluminum high-voltage bias foils where negative voltage is applied. In addition to the bias foils, there are two more grounded foils on the outermost surfaces over the outer bias foils. These grounded foils balance the electrostatic field on the bias foil and prevent the bias foil from deflecting towards the sense wires. They also provide a degree of safety by covering the bias foils with a grounded conductive shield. Two end plates hold the entire assembly together. The PWC assembly is filled with an 80% argon and 20% carbon dioxide gas mixture. Figure 3 has a more detailed view of the assembly. The bottom row shows the simulation model.

## BEAM-INSTRUMENTATION INTERACTIONS

The beam-detector interaction was simulated using G4Beamline [4]. The code incorporates all physical processes such as muon decay and particle-matter interactions. This study is restricted in the Delivery Ring and we assume one turn only.

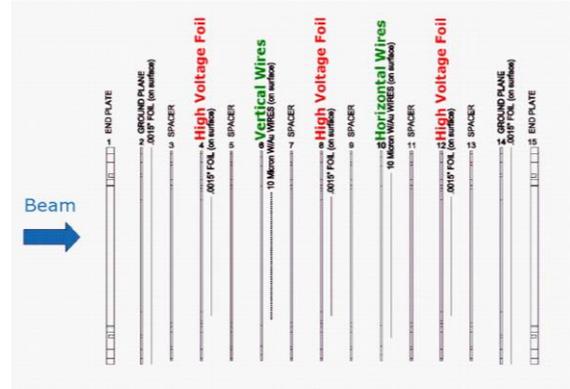

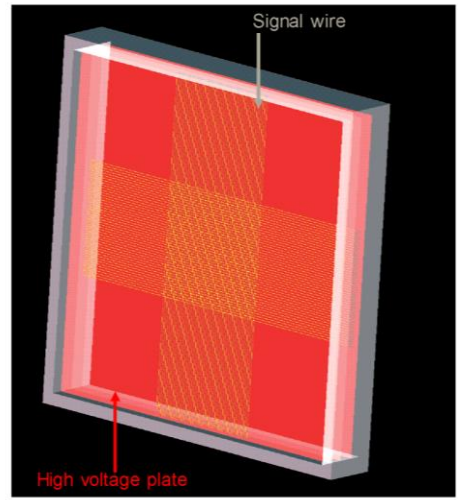

Figure 3: Schematic illustration of the PWC detector. Bottom figure shows the simulation model.

The DR is a rounded triangle and is divided into 6 sectors numbered 10-60. Each sector contains 19 quadrupoles and 11 dipoles. Other magnetic devices include correction dipoles and sextupoles. There are three straight sections – 10, 30, and 50, which are located directly beneath service buildings AP10, 30 and 50 respectively. The straight sections are regions of low dispersion while the arcs are dispersive regions. A typical cell in the arcs is comprised of an F-quadrupole with similarly oriented sextupoles on either side followed by a dipole or drift region, then a D-quadrupole also surrounded by sextupoles of the same convention and another dipole or drift region. Each straight section contains 2 PWCs resulting in 6 PWCs in total. Sections 10 and 20 contain one ion chamber each. For standard operations of the Muon (g-2) Experiment, the beam will loop the DR four times and consequently is expected to pass via these devices multiple times.

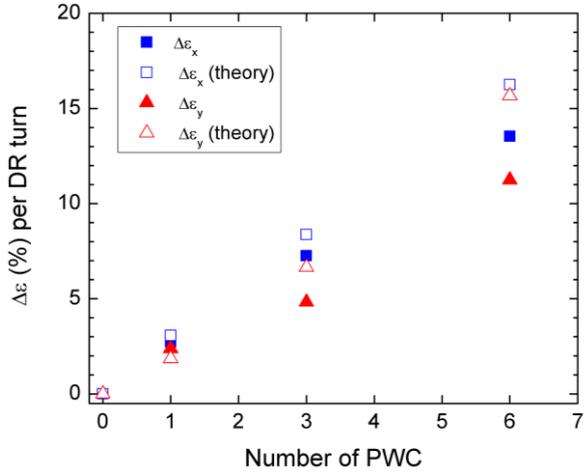

Figure 4: Horizontal and transverse emittance growth vs the number of PWCs along the beamline. Open symbols are results from theory while solid symbols are results from simulation.

Figure 4 displays the relative beam emittance growth as the muon beam loops the DR. The horizontal axis displays the number of PWCs along the beam path. Here we focus on the fourth turn only, while our initial distribution is the result of an end-to-end simulation described elsewhere [5]. We assume that all ion chambers have been retracted from the beamline. Figure 5 displays the relative transmission. Clearly, the PWCs can cause emittance growth which is accompanied by a substantial particle loss.

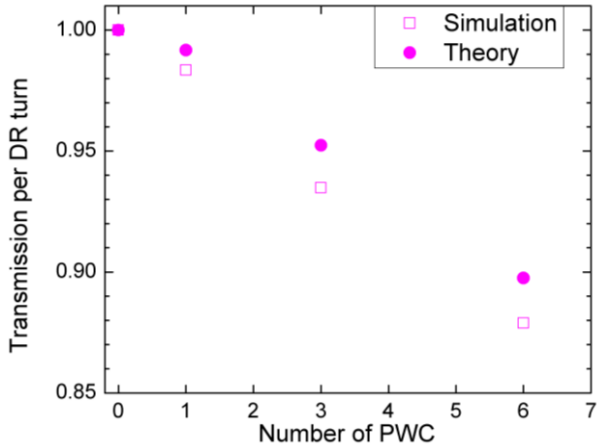

Figure 5: Transmission vs. number of PWCs. Open squares are simulation while solid cycles are results from theory.

Theoretically [6] we can estimate the emittance growth from multiple Coulomb scattering using the formula:

$$\varepsilon = \varepsilon_0 \sqrt{1 + \frac{\beta_0 \theta_{rms}^2}{(\varepsilon_0/\pi)}} \quad (1)$$

where $\varepsilon_0$ and $\varepsilon$ are the initial and final emittances after interaction, $\beta_0$ is the transverse beta function at the interaction point, and $\theta_{rms}$ is the rms angle due to multiple Coulomb scattering. Then, the particle loss, by assuming a Gaussian beam, can be estimated by the formula:

$$P_{loss} = e^{-A/2\varepsilon} - e^{-A/2\varepsilon_0} \quad (2)$$

where $A$ is the lattice acceptance which equals $40\pi$ mm.mrad for our case. Both theory and simulation show a near to 2% particle loss as the beam passes a PWC.

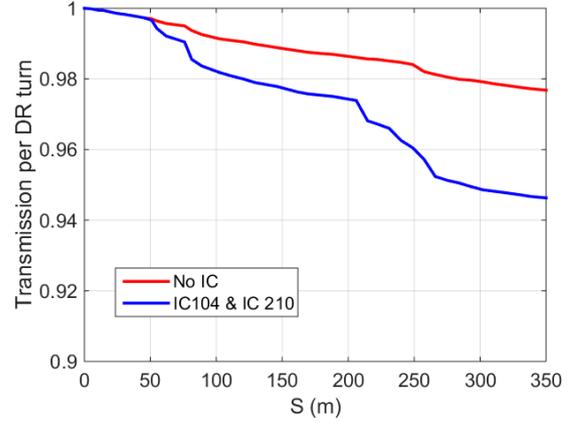

Figure 6: Simulated particle loss along one turn in the DR as the beam passes through the ion chambers.

Next, we examine the interaction of the beam with the ion chambers. In Fig. 6 we examine the number of particles vs. the distance for one turn in the DR. The two bumps are correlated to the location of the ion chambers. Our simulations show that both detectors contribute almost equally and result to a 2% loss of beam. The corresponding emittance growth in the horizontal and vertical planes is shown in Fig. 7.

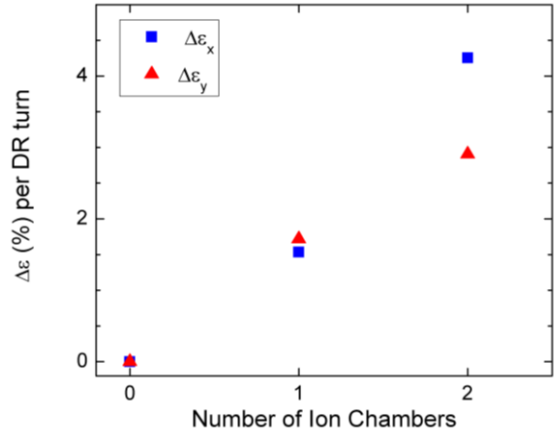

Figure 7: Emittance growth as a function of the number of ion chambers along the DR.